# GAS COUNTERS AGING FOR DUMMIES

Fabio Sauli[*]

European Organization for Nuclear Research (CERN)
1211 Geneva 23, Switzerland

ABSTRACT
Invited presentation at the 3rd International Conference on Detector Stability and Aging Phenomena in Gasous Detectors, this work summarizes the major outcomes of research on the processes of performance degradation of gaseous detectors on exposure to ionizing radiation, since the first observations with Multi-Wire Proportional Chambers in the early seventies of last century.

1. **Introduction**

   The invention in 1967 of the Multiwire Proportional Chamber (MWPC), with its outstanding tracking and rate capability compared to current devices, opened up new perspectives for particle physics experimentation [1]. The original MWPC design swiftly evolved to more complex structures with improved position accuracy, and capable to instrument large detection volumes: Drift Chambers, JET and Time Projection Chambers and others. The new detectors exploiting the process of charge multiplication around thin wires replaced the previous generation of gaseous counters, spark and streamer chambers, in experimental setups worldwide [2][3].

   Signs of performance degradation after exposure to even moderate radiation doses were observed rather early in the development of large size MWPCs [4]; Figure 1 shows the singles counting rate on a soft X-ray source as a function of applied voltage, at increasing radiation levels, before (full lines) and after (dashed lines) the exposure to around $10^7$ electrons per $cm^2$ from a radioactive source. As it can be seen, after the exposure the efficiency plateau has moved to higher values of voltage, and the onset of discharge to lower values; operation of the detector becomes soon unviable.

   Operating a damaged device at high gains permitted to observe the emission of visible photons around the regions affected by the degradation process, as shown in Figure 2, revealing the presence of localized or extended discharge clusters; bizarrely, the luminous emission clusters seemed to move around in sudden jumps, probably due to electrostatic repulsion. The cause of the emissions was identified to be the presence on the anode wires of dust particles and/or deposits of various nature, residuals of the manufacturing process or resulting from the creation under discharge conditions of solid or liquid polymers of the filling gas, generally using large fractions of hydrocarbons as quenchers.

---

[*] Corresponding author. email address fabio.sauli@cern.ch

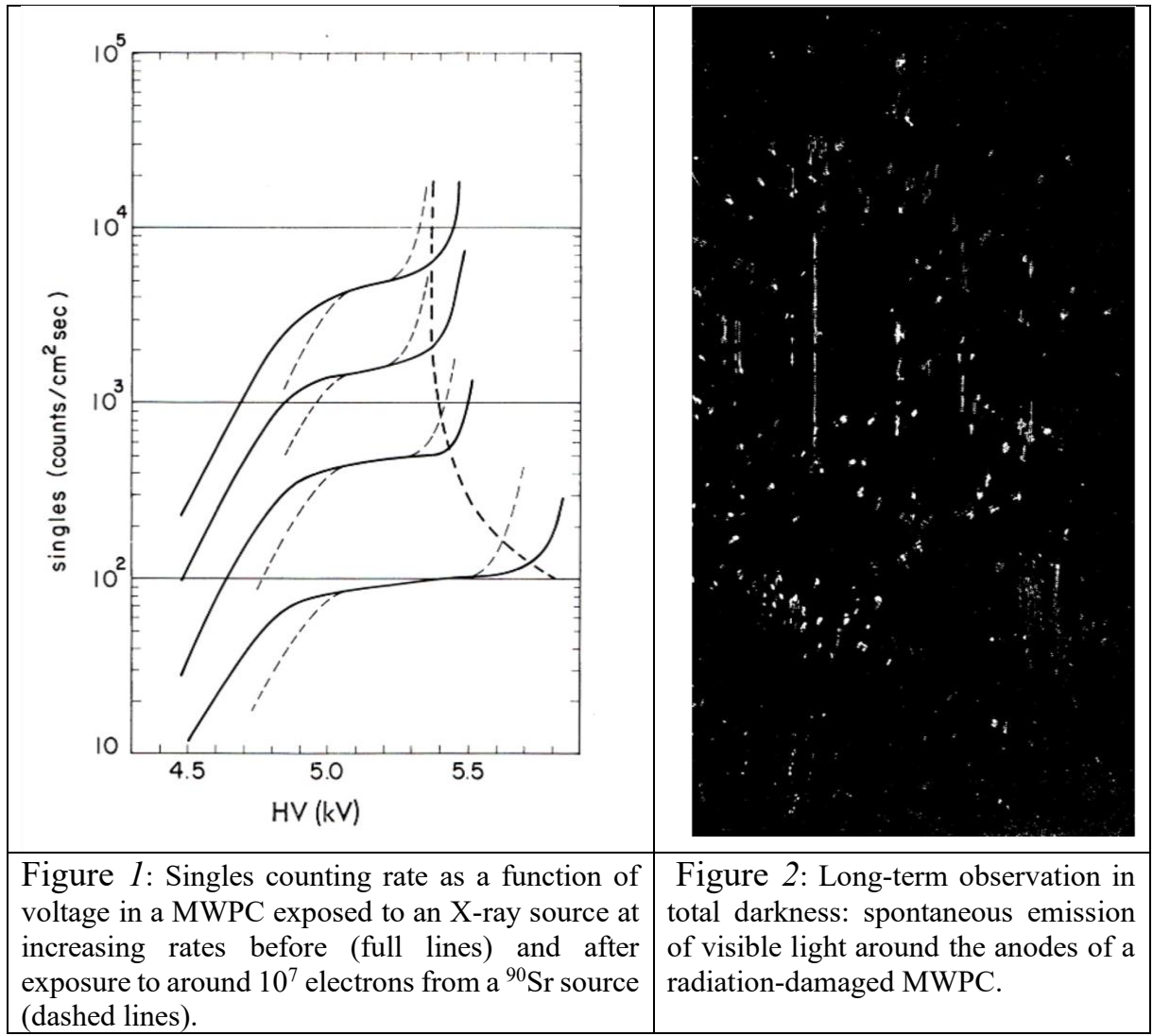

Figure *1*: Singles counting rate as a function of voltage in a MWPC exposed to an X-ray source at increasing rates before (full lines) and after exposure to around $10^7$ electrons from a $^{90}$Sr source (dashed lines).

Figure *2*: Long-term observation in total darkness: spontaneous emission of visible light around the anodes of a radiation-damaged MWPC.

In the early studies it was observed already that small changes in the gas input lines could largely enhance the degradation, or aging rate, of a wire detector, as seen in the examples of Figure 3 and Figure 4. In the first case a short portion of the original stainless steel tubing was replaced with a plastic PVC segment; in the second case, a mechanical flowmeter was inserted in the gas input line. Once initiated, the degradation persists albeit at a lower rate, indicating the presence of a permanent damage to the anodes. In the two examples, the culprits were recognized to be the release of volatile plasticizer composites in the PVC, and residuals of the lubricants used for manufacturing and operating the flowmeter.

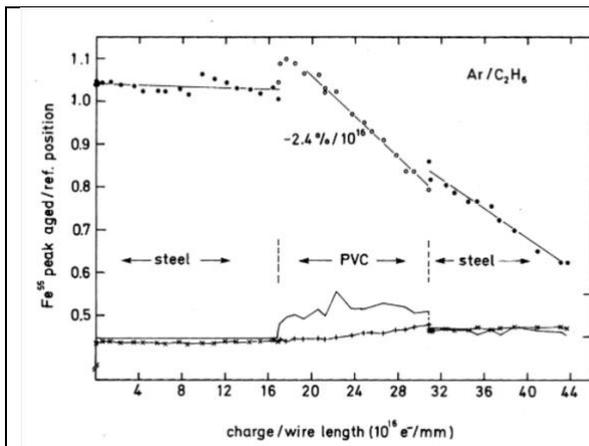 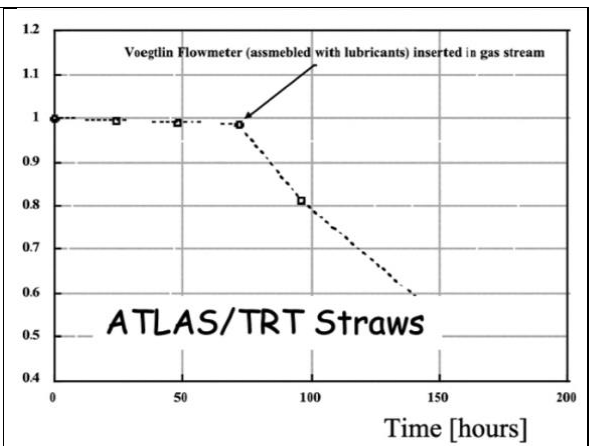

Figure *3*: Aging. effect of the replacement of a segment of steel with PVC tubing [5].

Figure *4*: Accelerated aging after insertion of a new flowmeter in the gas inlet [6].

Optical inspection of radiation-damaged detectors reveals the presence on the damaged anodes of heavy deposits, flakes and spikes, and discolorations on cathodes, see Figure 5. Anodic deposits are recognized to be carbon-based polymers of the filling gas, while cathode patterns seem the result of oxidation or reactive ions bombardment. For an extended discussion on the nature of the residues and their effect on the detector operation see the proceedings of the first Workshop on Radiation Damage to Wire Chambers held in Berkeley in 1986 [7] and review papers on the subject [8][9][10].

The residues commonly found on anode wires have been identified as the results of polymerization of the organic gases, used as quenchers to achieve stable operation: methane, ethane, isobutane and other hydrocarbons. Subject to collisions with energetic electrons in the avalanches, the molecules dissociate forming radicals, that recombine forming precursors that can attach to form long chains of compounds, as shown schematically in Figure 6 [11]. The processes are described by the field of plasma chemistry[1].

Particularly insidious oily deposits are often found on wires, either due to improper cleaning, or quickly growing under irradiation and due to the outgassing of silicon-based compounds used for tubing and manufacturing (lubricants, epoxies, sealants); an example was given in **Error! Reference source not found.**. The process of silicone polymers formation under avalanche conditions is particularly efficient, coating the electrodes with liquid or solid insulating layers, as discussed for example in Ref [12]. An updated list of compounds that can induce the formation of silicone polymers is given in Ref. [13], describing also various methods of recovering damaged detectors (see later).

---

[1] See the contribution to this conference by Katerina Kudnetzova.

The introduction in 1988 of the Micro-Strip Gas Chamber (MSGC) [14], extending the rate capability of wire-based detectors, generated new concerns about long-term survivability, as they appeared more prone to age in presence of even very small amounts of pollutants. Figure 7 shows an example of relative gain measured as a function of collected charge with an MSGC plate exposed to a soft X-ray flux and mounted either in a standard fiberglass-based frame or in a "clean" box made of stainless steel, aluminum and glass [15]; the difference in behavior is spectacular, indicating the extreme sensitivity of the detector to even minute amount of impurities.

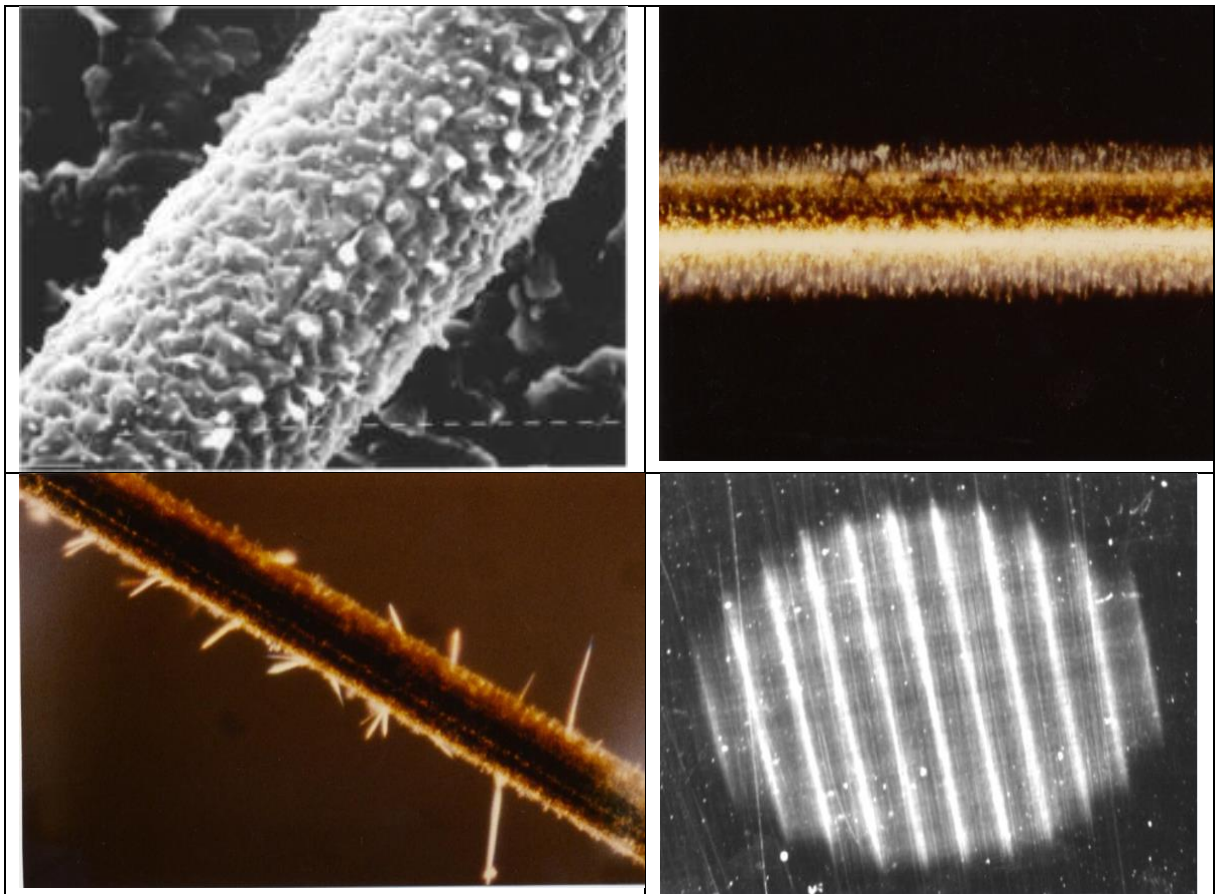

Figure 5: Examples of deposits on electrodes observed in aged wire chamber; the bottom right image is the patterned damage seen on a cathode plane facing the irradiated area.
Figure 5

## 2. Aging research projects

Confronted with the challenge to ensure medium- and long-term operation of the new family of devices, a cluster of laboratories launched a research project named RD-10 [16], approved at CERN in 1990, and continued with an growing list of collaborating institutes as RD-28 [17] and RD-51[18].

The core of the RD-10 project was a CERN-based laboratory aimed at ascertain the survivability of wire counters exposed to harsh radiation environments. Figure 8 shows schematically the setup used to study aging properties of detectors exposed to a continuous flux of ionizing radiation. A measurement of detector current as a function of time permits to record the radiation exposure, wile a monitor chamber exposed to an attenuated flux is used to correct variations in gas flow, temperature and pressure. To detect changes in performance, the proportional gain and resolution of the detector under test are recorded at regular intervals under an attenuated flux both inside and outside the irradiated area. A small but essential detail in the setup is the way the gas flow is vented, using a long exhaust pipe instead of the conventional oil bubbler, that was found to contribute to the silicone pollution despite being on the output line.

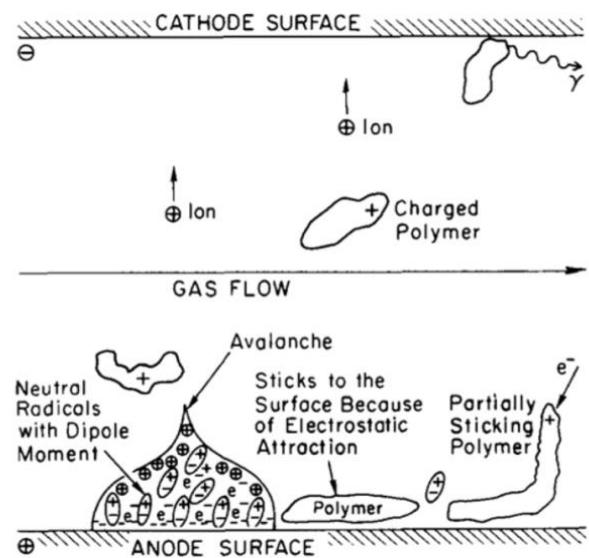

Figure 6: Schematics of the formation of polymers of organic molecules under avalanche conditions [11].

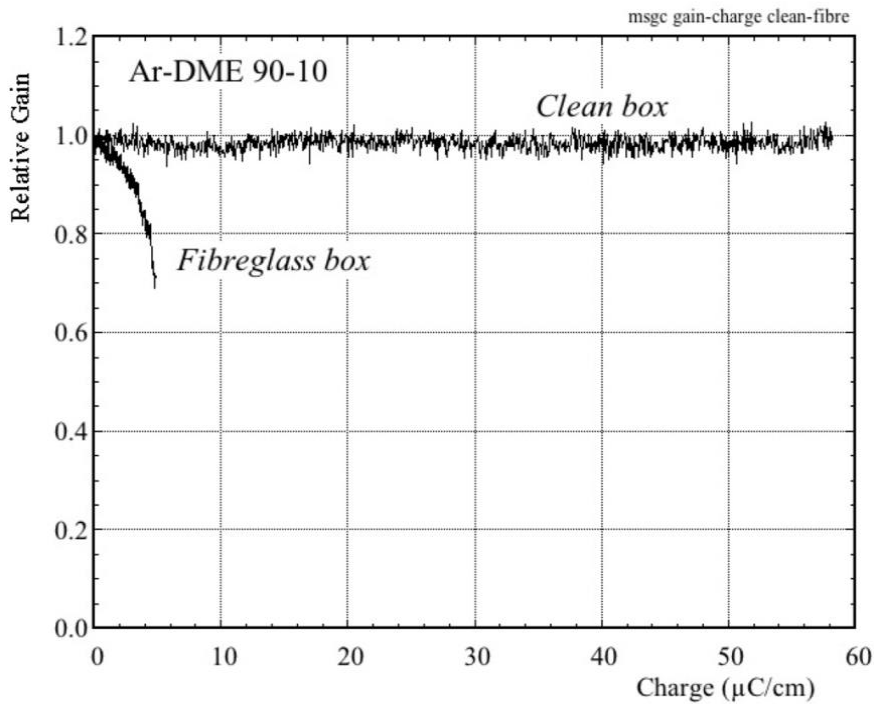

Figure 7: Relative gain dependence from collected charge measured exposing to ionizing radiation an MSGC assembled within a fiberglass-epoxy container and in a "clean" stainless stees and aluminum box [15].

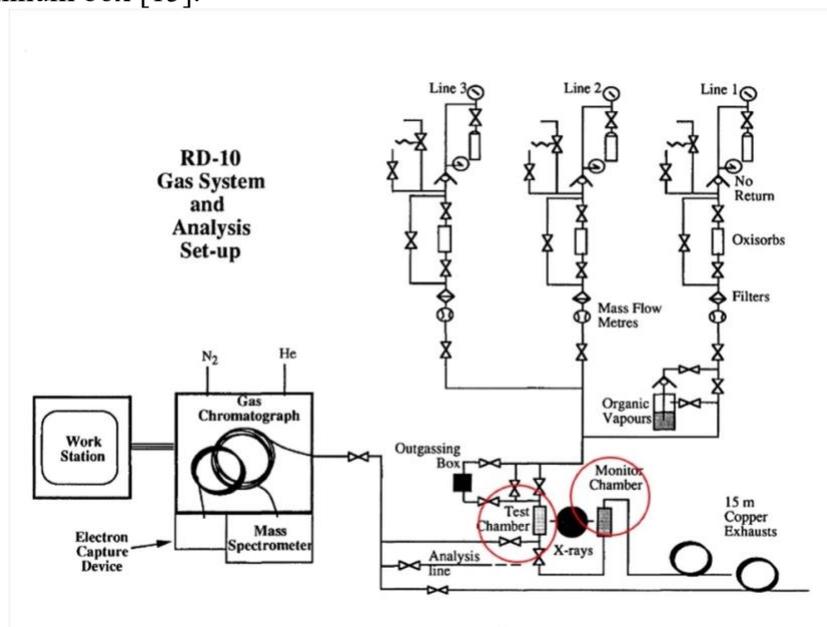

Figure 8: The RD-10 setup with the chamber under test irradiated with a high X-ray radiation flux.. The volatile yields carried with the gas flow are analyzed with a gas chromatograph.

The natural unit to express the survivability of gaseous counters is the amount of collected charge per unit surface; for MWPCs, where the distance between wires modulates the charge collection, it is customary to express the charge per length of wire. Other units are used in the literature, making often difficult to compare results: time of exposure, count per unit surface, years of operation. It should be noted that aging tests are necessarily accelerated, often by a large factor, raising doubts about the pertinency of the results in normal operating conditions: it is indeed a know outcome of plasma chemistry that the polymerization rate depends on charge density, roughly increasing with the square root of the dose. The effect is illustrated in Figure 9, a collection of aging rate measurements in counters with Ar-$CH_4$ fillings, as a function of irradiation charge density [19]. Although compiled for different counters and conditions, the result is indicative of the general trend of high-current measurements to be optimistic, often by a large factor.

The research permitted to find manufacturing materials, gases and operating conditions suitable to extend the detectors lifetime, and to identify potentially dangerous components; the results are summarized in extended tables included in Refs [19] and [10]. The list of components to unconditionally avoid includes: plasticizers (Phthalates), sealants (RTV), vacuum greases, oil bubblers, silicone oils, Duo-Seal, some epoxies (G10) and many other materials.

A parallel line of research has been conducted by a group at Helsinki University, operating a laboratory setup similar to the CERN system; an example of gas chromatograph analysis of stable organic compounds created by electron avalanches in a proportional counter operated with argon-methane is shown in Figure 10 [20]; many of the species identified are potentially able to produce stable polymers.

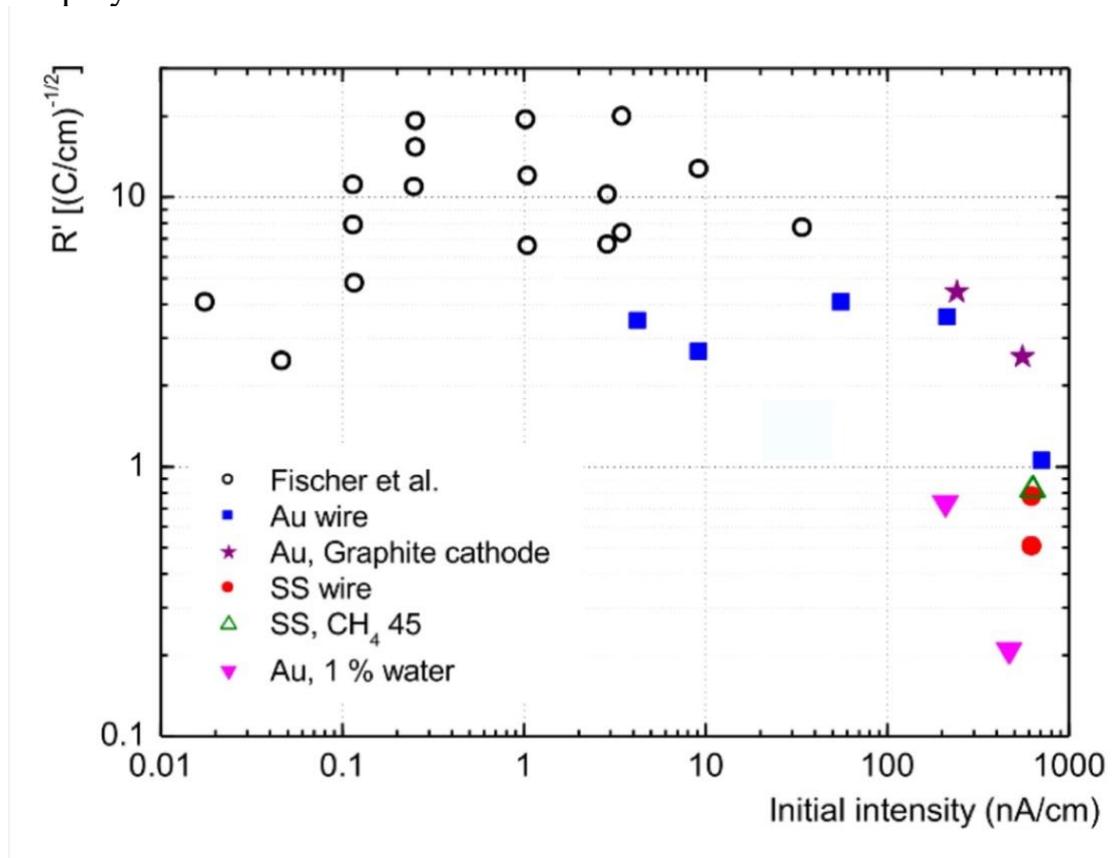

Figure 9: Compilation of aging rate measurements in different wire counters as a function of irradiation current density [19].

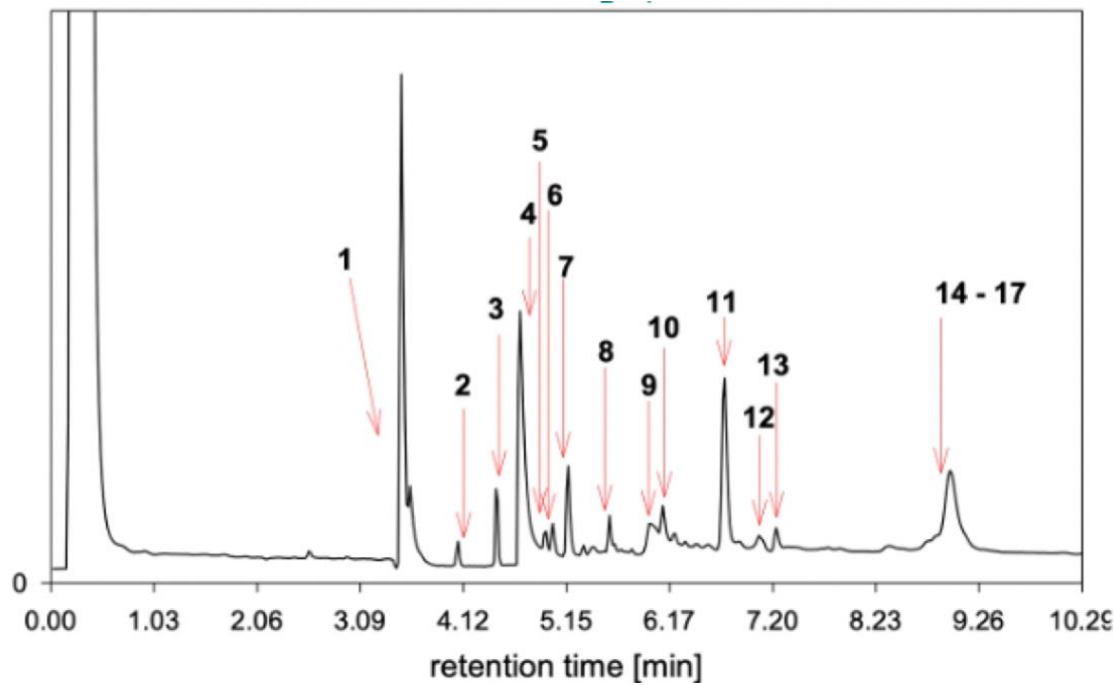

Figure 10: Gas chromatograph spectrum of stable organic compounds created by electron avalanches in an argon-methane. Close to half of the species are able to polymerize. For the identification of the peaks see Ref [20].

## 3. Etching of deposits and detector recovery

Carbon tetrafluoride ($CF_4$) has been used as additive in proportional counters owing to its high electron drift velocity and low diffusion, consenting to improve the time resolution and localization accuracy. It soon appeared that the formation in the avalanches of fluorinated radicals could help as etching agent for organic and silicone compounds created in the processes described above. Not only the addition of $CF_4$ prevents the formation of deposits, but it has been used to recover wire chambers affected by the polymerization process.

Figure 11 is an example of recovery of a section of a MWPC damaged by irradiation in an argon-ethane gas mixture, after replacement of the hydrocarbon with a carbon fluoride-isobutane gas filling [21]. Figure 12 is another example of complete recovery of an aged wire after exposure to a soft X-ray source in a $CF_4$ mixture [13]; this reference describes also a "training" procedure, consisting in repeated exposures to a radioactive source in the mixture.

While polymer and silicone deposits on anode wires result in a local reduction of gain, thin insulating layers deposited on cathodes induce a spontaneous electron emission through the so-called Malter effect, schematically represented in Figure 13. As these secondary electrons return to the anodes and release more ions in the avalanches, the process may become divergent and result in sustained discharges.

A training procedure, similar to the one described for anodes, has been used to recover the LHCb muon chambers, and is illustrated in Figure 14: it consists in repeated powering cycles after adding oxygen to the standard operating gas mixture (Ar-$CO_2$-$CF_4$) [22].

It should be noted that carbon tetrafluoride, despite its benefits in curing and preventing polymerizations, is a greenhouse gas, difficult to procure and rather aggressive for other manufacturing components of the detectors.

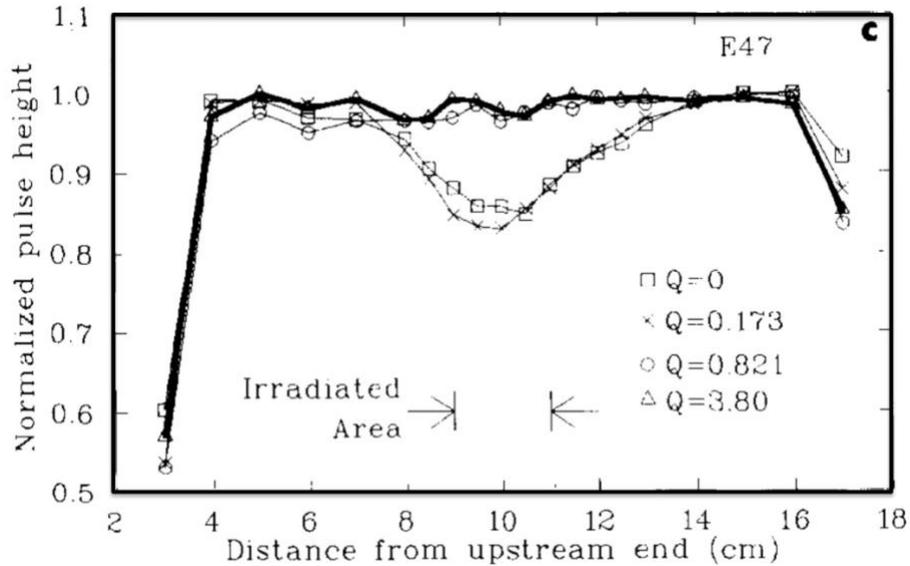

Figure *11*: Recovery of an aged section of a MWPC filled with a $CF_4$-$iC_4H_{10}$ mixture [21].

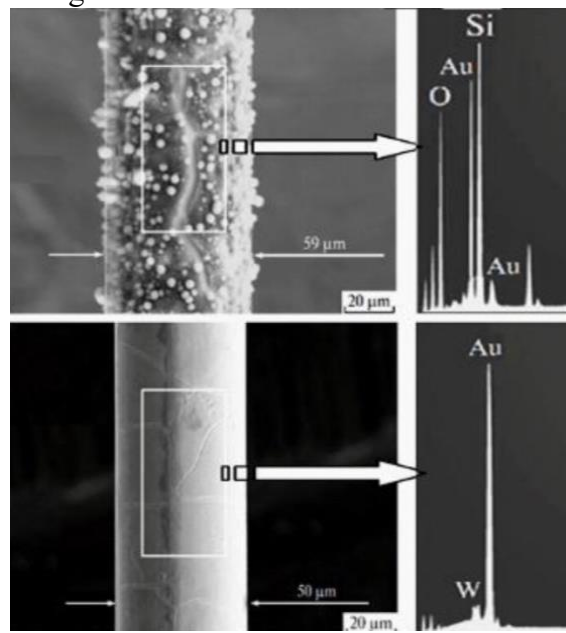

Figure *12*: Recovery of an aged wire with exposure to a radiation in a gas mixture containing 10% $CF_4$ [13].

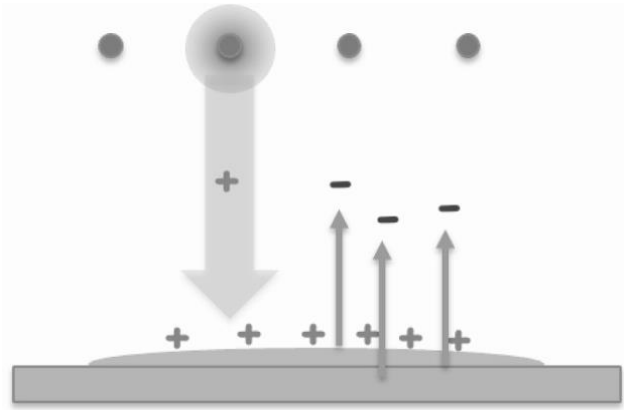

Figure 13: Malter effect: spontaneous secondary electron emission due to the high dipole field created by positive ions accumulating on a thin insulating layer [23].

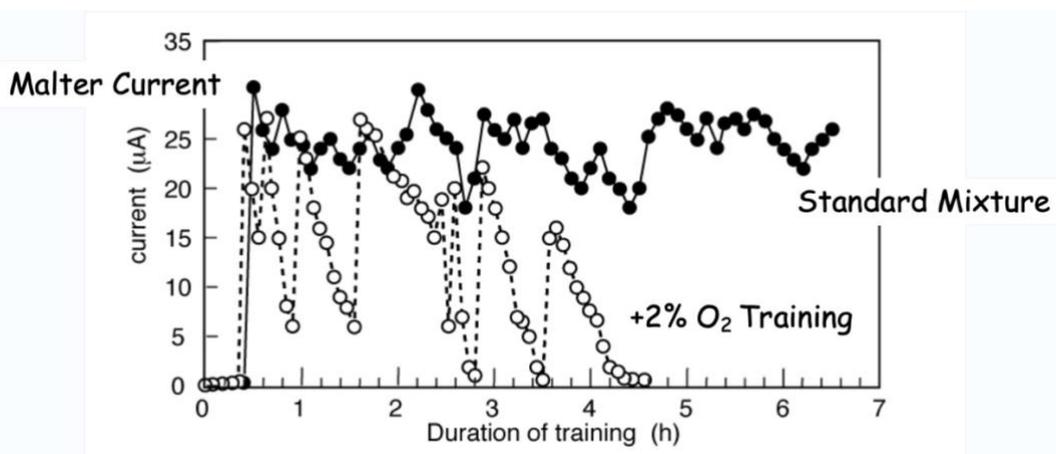

Figure 14: Training process to reduce the Malter current with addition of 2% oxygen in the gas mixture [22].

The experience accumulated in the study study of aging processes in wire chambers resulted in a set of "Golden Rules for Aging Prevention", here summarized:
- Use of ultra-pure gases
- Use of non-organic quenchers ($CO_2$)
- Choice of certified non-outgassing construction materials
- Use of non-polymerizing additives: methylal, alcohol
- Improved cleaning protocols for all materials
- Avoidance of silicon-containing components: tubing, epoxies, sealants

Aged detectors can be partially or totally recovered with various "cleaning" procedures:
- Training: cycles of sustained exposure to radiation with gas components having etching properties ($CF_4$, $O_2$)
- Zapping: burning the deposits with high current on anodes (rarely possible)

As indicated in the following section, the aging problem has been partly if not completely solved by the introduction, in the late nineties, of the new family of wireless devices collectively named Micro-Pattern Gaseous Detectors (MPGD).

## 4. Micro-Pattern Gaseous Detectors

An important outcome of the aging studies is that the formation of organic and silicone polymers is largely enhanced in high electric fields, an outcome also of basic plasma chemistry [11]. Both in MWPCs and MSGCs the field at the anode surface is very high, typically around 500 kV/cm, enhancing the dissociation of gas molecules in reactive species. The situation is rather different for the new families of detectors, and in particular the MICROMEGAS [24] and the Gas Electron Multiplier (GEM) [25] where avalanche amplification occurs in much lower fields, typically 20-30 kV/cm. Not only this considerably reduces (even though not eliminates) the polymers formation, but also the presence of deposits on electrodes has modest effect on the multiplication mechanism, that occurs in the gap or in the holes of the two devices.

As the radiation resistance of MPGDs is the main subject of this conference, covered by many contributions to the proceedings, only a brief summary of relevant previous observations will be presented here.

Adopted for the upgrades of several LHC experimental setups, MICROMEGAS and GEM detectors have been subject of extensive exposures to radiation, with continuous monitoring of their main performances (gain and energy resolution). Figure 15 is the result of a long-term irradiation with an X-ray generator of a MICROMEGAS prototype of the ATLAS muon upgrade; operated with an argon-$CO_2$ mixture, the detector gain remains unaffected up to a collected charge of 230 mC/cm [26]. Figure 16 is a similar measurement on a Triple-GEM prototype for the CMS forward muon upgrade; the total collected charge on exposure to an X-ray generator is 900 mC/cm$^2$, without noticeable changes in gain and resolution [27]. In both cases, the results ensure the survivability of the detectors in the LHC High-Luminosity long-term operation [2].

---

[2] Contributions to this conference report an increase of these limits to larger values of collected charge.

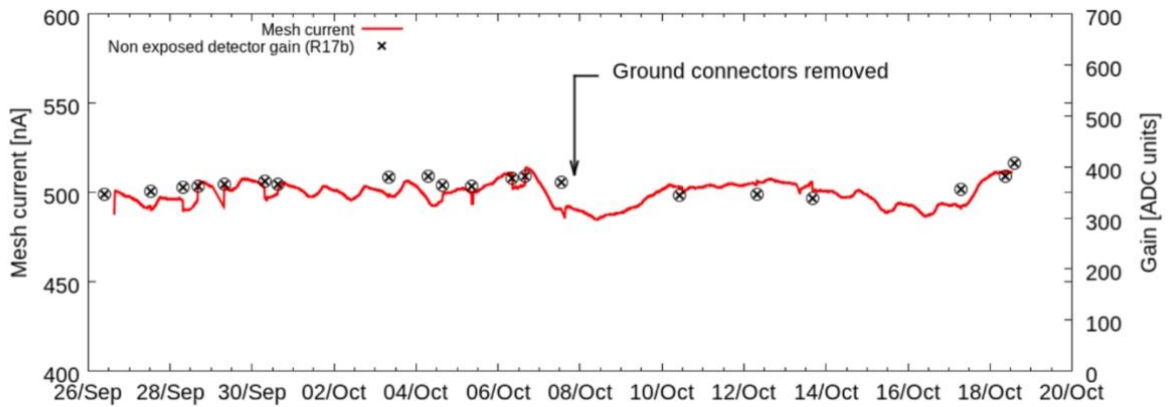

Figure 15: Long-term gain measurement of the ATLAS MICROMEGAS prototype. The total collected charge is ~230 mC/cm$^2$ [26].

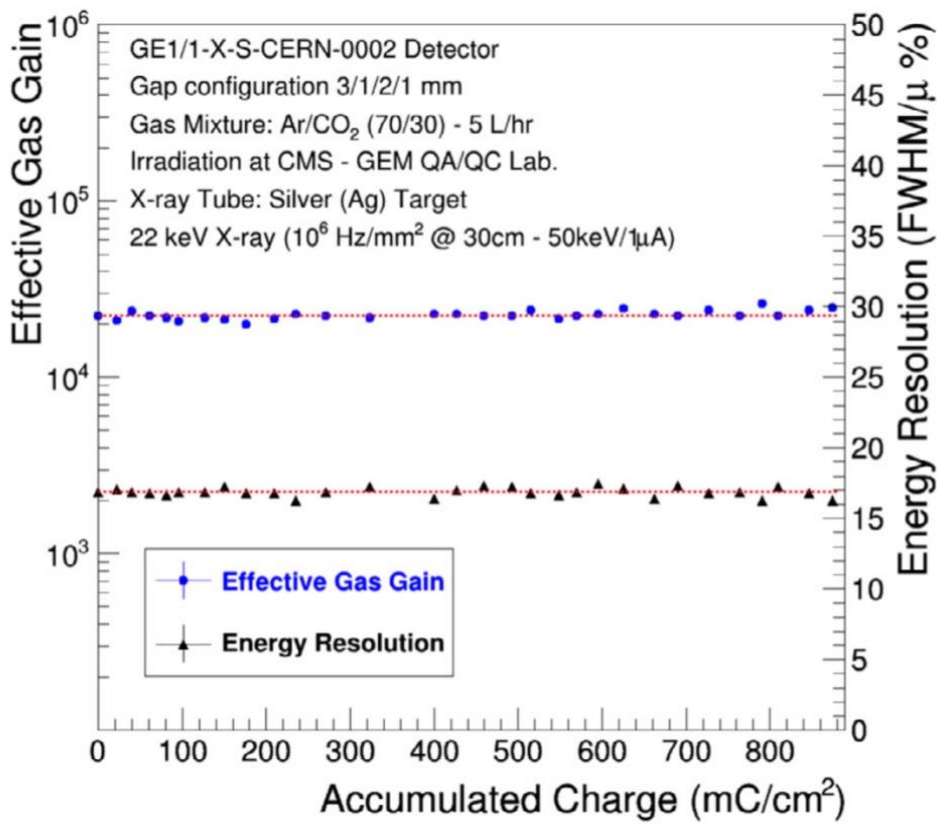

Figure 16: Long-term gain and resolution of the Triple-GEM prototype for the CMS forward muon upgrade [27].

Aging has been observed in MPGDs when the "golden rules" have not been followed. As an example, in a prototype of the ALICE GEM-TPC, a gain reduction by ~30% has been observed in Ar-CH$_4$ after less than 100 mC/cm$^2$ of collected charge; the performance recovers and remains constant replacing the organic quencher with $CO_2$, Figure 17 [28].

An even more manifest result of an unfortunate choice of components has been reported for the COMPASS Pixel Triple-GEM: a deterioration in the area exposed to the beam and the presence of silicone and sulfur deposits on electrodes
12

after an accumulated charge of less than 200 mC/cm$^2$, Figure 18. The culprit has been identified to be the use of a sealant (DowCorning 1-2577) to fix a leak.

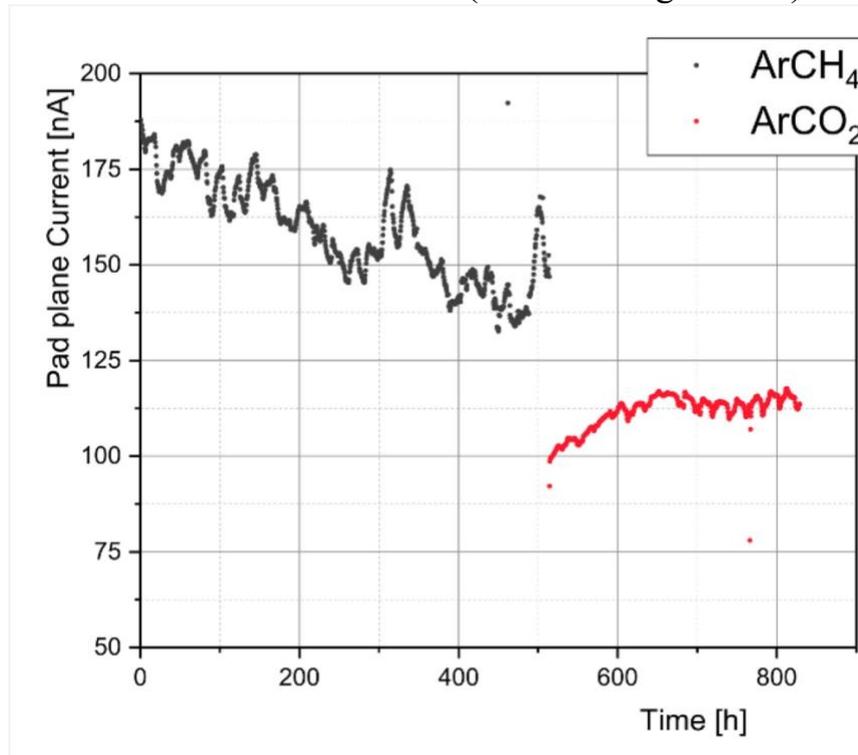

Figure 17: Gain decrease in an ALICE GEM-TPC prototype operated with Ar-CH$_4$ (left points), and recovery in Ar-CO$_2$ (right points) [28].

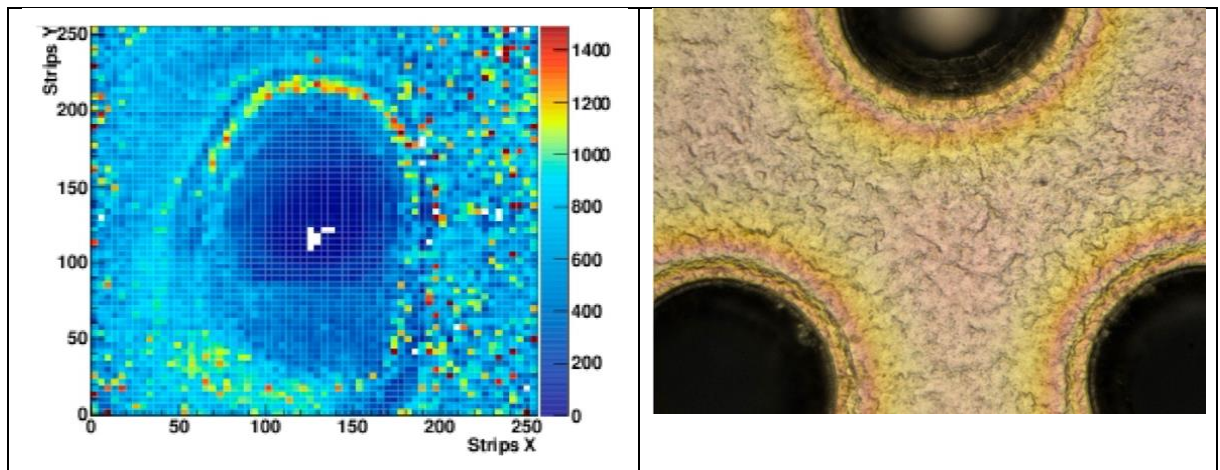

Figure 18: Gain decrease (left) and damages of the electrodes in the irradiated area (left) due to the pollution of a sealant in the COMPASS Pixel GEM detector [29].

## 5. Conclusions and summary

Fifty years after the discovery of fast loss of performances of gaseous detectors exposed to high radiation levels, an extensive research has helped identifying the factors responsible of the degradation. A set of "Golden Rules" apt both at preventing aging and curing damaged detectors helped operating large detector systems in harsh conditions. The emergence in the late nineties of new families of devices, collectively named Micro-Pattern Gaseous Detectors, less prone to the aging processes, holds promises to ensure correct long-term operation in increasingly more hostile environments as those encountered at the High Luminosity colliders and in other applications.